\documentclass[preprint]{aastex}

\shorttitle{IR Observations of Nova Muscae 1991}
\shortauthors{Gelino et al.}

\begin{document}

\title{Infrared Observations of Nova Muscae 1991: \\
Black Hole Mass Determination from Ellipsoidal Variations}

\author{Dawn M. Gelino,\altaffilmark{1,2} \email{dleeber@nmsu.edu} Thomas E. 
Harrison,\altaffilmark{2} \email{tharriso@nmsu.edu} \& Bernard J. McNamara \email{bmcnamar@nmsu.edu}}
\affil{Department of Astronomy, \\
New Mexico State University, Las Cruces, NM 88003}

\altaffiltext{1}{formerly Dawn M. Leeber}
\altaffiltext{2}{Visiting Astronomer, Cerro Tololo Inter-American Observatory, 
National Optical Astronomy Observatory, which is operated by the Association of 
Universities for Research in Astronomy, Inc. (AURA) under cooperative agreement 
with the National Science Foundation.}

\begin{abstract}
We have obtained infrared photometry for the soft x-ray transient GU Mus.
We present $J$ and $K_s$ band light curves modeled with WD98, the newest version
of the Wilson-Devinney light curve modeling code.  Using detailed 
models for the expected ellipsoidal variations due to the non-spherical 
secondary star, we show that the most likely value for the orbital
 inclination is 54$\pm$1.5$^o$.  This inclination angle is 
consistent with those previously published, but has a much smaller 
error.  This inclination implies a primary black hole mass of 6.95$\pm$0.6 
M$_{\odot}$.  While we do not see any evidence for contamination of our infrared light curves from other sources in the system, a conservative model with a contamination level of 15\% increases the uncertainty in the inclination angle to 54$^o$$^{+4}_{-1.5}$.  

\end{abstract}

\keywords{binaries: close
$-$ stars: black holes
$-$ stars: individual (Nova Muscae 1991)
$-$ stars: low mass
$-$ stars: variables: other}

\section{Introduction}

Soft X-Ray Transients (SXTs) are binary systems believed to contain a black
hole primary and low mass secondary star.  They display 
large and sudden x-ray and optical outbursts that are believed 
to be the result of a sudden, dramatic increase in the mass accretion rate 
onto the compact object.  In quiescence, SXTs are very faint at x-ray 
and optical wavelengths, however, in this state, the secondary stars 
can dominate the system luminosity.  

GU Mus (=Nova Muscae 1991; XN Mus 91; GRS 1124-68) 
was discovered by the Ginga and Granat satellites on January 9, 1991 
(Makino 1991; Lund 1991).
Since then, it has been studied by Remillard et al. (1992), Orosz et al. 
(1996; hereafter OBMR), Casares et al. (1997; hereafter CMCMR), Shahbaz, Naylor, 
\& Charles (1997; hereafter SNC), and references therein.  
These authors have found an orbital period of 10.38 hours, a secondary star 
spectral type of K4V, and a secondary star radial velocity semi-amplitude of 406$\pm$7 km s$^{-1}$.
  GU Mus has an implied mass function, the
minimum mass of the compact primary object, of 3.01$\pm$0.15 M$_{\odot}$ 
(OBMR).  This mass function combined with the radial 
velocity curve of the secondary star, established the primary object as 
a dynamical black hole.  Both in outburst and quiescence, GU Mus closely 
resembles the prototype black hole SXT, V616 Mon (A0620-00).

In quiescence, most SXT's light curves reveal ellipsoidal variations ascribed
 to the secondary star.  Since the secondary star fills its Roche lobe, the 
surface 
area seen by an observer on Earth changes as the star orbits the compact 
object.  This changing line-of-sight surface area corresponds to a changing 
apparent brightness.  The amplitude of the ellipsoidal variations is 
determined by 
the orbital inclination angle of the system.  By combining the orbital 
inclination angle with the observed mass function, the SXT system parameters 
can be determined.  To this end, we have obtained new infrared observations of
 GU Mus and model its ellipsoidal light curves using a sophisticated light 
curve modeling program.

Other authors have also attempted to model observed light curves of GU Mus and 
other SXTs. In the case of GU Mus all but one of these papers, SNC, have 
observed the SXT in the optical.  If one is searching 
for the purest ellipsoidal variations, one should observe at a wavelength where
the secondary star provides the majority of the system luminosity.  In the 
optical, the accretion disk and hot
spot can contaminate, if not dominate, the system luminosity.  Even in 
quiescence, weak and variable accretion might be taking place, and 
there could be a modest amount of dilution of the optical light curve by 
the accretion disk, hot spot, and possible heating of the secondary star 
(SNC).  On the other hand, in SXTs with K and M type companions, the 
secondary star can dominate the quiescent binary's luminosity in the 
near-infrared.  For GU Mus, CMCMR find that the secondary 
star supplies between 85 - 88\% of the observed flux in the $R$ band. 
This percentage will be even higher at longer wavelengths.  Therefore, 
observations in the infrared will reveal more genuine
 ellipsoidal variations than observations in the optical.  

The only way to obtain the orbital inclination angle in non-eclipsing binary 
systems is from modeling.  Previously published inclination angles for GU Mus 
range from 39$^o$ (lower limit from Antokhina \& Cherepashchuk 1993) all the 
way to 83$^o$ (upper limit from Hua \& Lingenfelter 1993).  More recently, the
 inclination has been refined to 54$^o$$^{+20}_{-15}$ by SNC, and 
60$^o$$^{+5}_{-6}$ by OBMR.  These translate into primary mass ranges of 3.8 
M$_{\odot}$ $-$ 10.5 M$_{\odot}$, and 5 M$_{\odot}$ $-$ 7.5 M$_{\odot}$, 
respectively.  The rather large ranges in the inclination angles and primary 
masses that have been published for GU Mus and other SXT systems, has 
prompted us to determine 
whether we could more precisely determine the primary mass of the 
black holes in these SXTs, by using the most recent version of the 
Wilson-Devinney light 
curve modeling program to model new infrared observations.  We have observed 
five SXTs in the infrared, but have chosen to present results on GU Mus first 
out of our sample, since it is the simplest of these systems to model.  With 
no hot spot and very low accretion rate, we probably have the best case 
scenario for modeling relatively uncontaminated ellipsoidal variations.

In section 2, we describe our observations and data reduction process, as well 
as present our infrared photometric light curves. Section 3 describes our 
infrared light curve modeling procedure using WD98, the newest 
version of the Wilson-Devinney light curve modeling code. We provide
details of how we chose the relevant input parameters, present the 
resulting models at $J$, and $K_{s}$, and discuss the error in the orbital 
inclination angle. Finally, Section 4 discusses the 
implications of the models, and compares our results to those previously 
published.

\section{Observations \& Data Reduction}

GU Mus was observed using OSIRIS (see Pogge et al. 1999) on the Victor M. 
Blanco 4-m telescope at the Cerro Tololo Inter-American Observatory on 2000 
February 20 and 21.  On February 20, GU Mus was observed from 1:06 to 
9:08 UT with the camera at the f/7 plate scale (0.161"/pixel). On 21 
February from 2:16 to 9:12 UT we observed GU Mus with the camera at the 
f/2.8 plate scale
(0.403"/pixel) due to poorer seeing conditions.  With an 
orbital period of about 10.38 hours, the first 
observing session covered two-thirds of an orbital period, while the second 
covered three-quarters of an orbital period.  Photometric data were obtained 
in the OSIRIS $J$ ($\lambda_c$=1.215 $\micron$) and $K_{s}$ ($\lambda_c$=2.157 
$\micron$) filters. Our observing sequence consisted of a single $J$ image at 
one position, a beam switch, two additional $J$ images, a beam switch back to 
the original position, and then one more $J$ image (an ABBA sequence).  We then 
switched to the 
$K_s$ filter and repeated the procedure.  Each individual f/7 (f/2.8) $J$ 
image consisted of 1 frame of 180 (75) seconds, while the corresponding f/7 
(f/2.8) $K_{s}$ images consisted of 2 (10) coadded frames of 35 (4) seconds 
each.  
	
Dome flats at both plate scales were obtained.  Before processing, all of 
the data were linearized using the {\it irlincor} package in IRAF with the 
coefficients supplied in the OSIRIS User's Manual (Pogge et al. 1999).  
After averaging the two images at one position, we subtracted them
from the average of the two images at the other position.  These sky, and 
bias-subtracted images were then flat fielded using the appropriate dome flat.
Figure 1 shows a $J$ band image of GU Mus and the five nearby field stars used 
for the differential photometry.  

Aperture photometry was performed on GU Mus and the five nearby field stars 
shown in Figure 1.  
Using the IRAF {\it phot} package, a differential light curve in both $J$ and 
$K_{s}$ was generated with each f/7 point being the average of the two beam 
switched images, and each f/2.8 point being the average of four beam 
switched images.  Our differential photometric results show that over the 
course of our observations, the comparison stars did not vary more than 
expected
from photon statistics.  The final $J$ and $K_{s}$ differential light 
curves of GU Mus, phased to the SNC ephemeris, are presented in Figure 2. 

Optical observations of GU Mus were taken with the Cassegrain Focus CCD Imager on the 0.9 meter telescope at CTIO on 2001 March 14 at 3:40 UT.  Data were obtained in the $V$, $R$, and $I$ bandpasses for the purpose of determining the quiescent optical colors of the system.  The 600 second exposures were zero corrected and flat fielded before aperture photometry was performed.  Landolt standard fields were also observed in these optical bands, and were used to correct the apparent magnitudes of GU Mus.  The resulting colors can be found in Table 1, along with previously published optical colors for this sytem.  

\section {Modeling the Light Curve of GU Mus}

Before we can begin modeling the light curves of GU Mus, we have to determine 
a large number of parameters.  Some of these are intrinsic to the binary system
 (Sect. 3.1), while others arise from the study of stellar atmospheres (Sects. 
3.3 \& 3.4).

\subsection{System Parameters}

The most important input parameter is the nature of the secondary star.  In 
1996, OBMR compared their observed spectrum of GU Mus to the 
spectra of various comparison stars, and determined a spectral type of K4V$\pm$1
 for the secondary star.  In 1997, CMCMR found that the best match to
 their H$\alpha$ spectrum was that of a K3V - K4V.  Both sets of authors state 
that the secondary star in GU Mus must be slightly evolved since a `normal' 
K4V would underfill its Roche lobe.  The secondary can not be a giant, 
however, since then it would not fit into an orbit with the observed orbital 
period.  

Knowing the secondary's spectral type does not mean that the mass of the 
secondary is also known.  These are most likely not `normal' stars, as they 
have survived a supernova explosion and are losing mass. Also as just noted, 
GU Mus probably has a slightly evolved secondary. Determining the mass of the 
secondary is difficult, and provides a large uncertainty to our derivation of 
the 
primary black hole mass.  Even though there is no definitive way to measure the mass 
of the secondary, there are techniques to determine the mass ratio, $q$, for 
the system.  Orosz et al. (1994) analyzed Balmer
 emission lines to find K$_1$ to use as an indicator of the motion of the 
primary.  They then compared K$_1$ to their K$_2$ to obtain a $q$ = K$_1$/K$_2$ =
 0.133$\pm$0.019.  Since their K$_1$ probably arose in the accretion disk, it is 
only a limit to the true radial velocity of the primary.
  CMCMR calculated a system mass ratio of $q$ = 0.128$\pm$0.04 from combining their 
$v$ sin$i$ value for the secondary star with geometrical constraints.  They 
were surprised to find that their mass ratio was close to the value found by 
Orosz et al. (1994). Fortunately, the orbital inclination of 
the system is relatively insensitive to the mass ratio (SNC). We will return 
to this point in section 3.6.

Interestingly enough, CMCMR did not detect a hot spot 
in their H$\alpha$ Doppler map of GU Mus.  This indicates that the H$\alpha$ 
luminosity of the hot spot is smaller than, or equal to, that of the 
accretion disk. CMCMR did detect some emission from the accretion disk; 
however, and suggest that, ``Perhaps the mass transfer rate in N. Mus is not sufficient to supply the luminosity of the hot-spot.''

\subsection{Model Setup}

To model the infrared light curves, we have used WD98, the newest version of 
the Wilson-Devinney light curve program (J. Kallrath 1999, private communication; R. E. Wilson 1999, private communication).  
WD98 is an enhanced version of WD95 (Kallrath et al. 1998), updated 
with new features such as the addition of semi-transparent circumstellar 
clouds, a simple spectral line profile capability for fast-rotating 
stars, an option to work with either observed times or phases,
and the conversion of all the variables to double precision.  Some
of the relevant features of WD98 include: Kurucz atmosphere models 
for numerous wavelengths, a choice of three different limb darkening 
laws which are sensitive to changes in gravity, proximity and eclipse effects,
 the option for hot or cold
stellar spots, and several different modes of operation for various 
system geometries.  Descriptions of the earlier versions of the program can be 
found in papers by Wilson and Devinney (1971) and by Wilson (1979, 1990, 1993).
A recent application of WD95 can be found in Milone et al. (2000). 

Briefly, WD98 works as follows. It takes the photospheres of the stars
and divides them up into a multitude of surface elements. The amount 
of light coming from each element is calculated based on the binary 
system input parameters.  All of these surface elements are then 
summed together, corrected for the line-of-sight geometry, to create the 
final light curve.

There are a number of input parameters needed to generate a model light curve 
for an SXT such as GU Mus. We discuss each of the most important of these 
parameters in the following subsections.  We have made use of the best available 
system parameters. We list the most important wavelength-independent input 
values to WD98 in Table 2, and the wavelength-dependent parameters in Table 3. 
Units are shown where appropriate.  We ran WD98 in a mode 
set up to produce a model for a semi-detached binary with the secondary
star automatically filling its Roche lobe (Mode = 5), and the primary having 
such a large gravitational potential, that it essentially has a radius of zero.  
We now discuss some of the input parameters in detail.

\subsection{Limb Darkening}

The most important parameter that affects both the shape and the 
amplitude of the ellipsoidal variations is limb darkening.  WD98 
allows you to choose from several different forms of limb darkening: 
linear, logarithmic, or square-root.  The linear law,
$$ I_{\lambda}(\mu) = I(1)(1 - x_{\lambda}(1 - \mu)), $$
was first investigated by Milne in 1921. In this equation, $I_{\lambda}$
is the beam intensity at wavelength $\lambda$, $\mu$ is the cosine 
of the angle between the atmosphere normal and the beam direction, 
and $x_{\lambda}$ is the limb darkening coefficient.  As an alternative 
to this, Klinglesmith \& Sobieski (1970) proposed the logarithmic law,
$$ I_{\lambda}(\mu) = I(1)(1 - x_{\lambda}(1 - \mu) - y_{\lambda} 
\mu ln(\mu))), $$ where $y_{\lambda}$ is the non-linear limb-darkening 
coefficient. Di\'az-Cordov\'es \& Gim\`enez (1992) introduced the
square-root law,
$$ I_{\lambda}(\mu) = I(1)(1 - x_{\lambda}(1 - \mu) - y_{\lambda} (1 -
\sqrt{\mu})). $$ When compared to ATLAS atmosphere models, the logarithmic 
law appears to fit UV models the best, while the square-root law 
appears better at infrared wavelengths (Van Hamme 1993).  Models run 
by Claret (1998) for very low mass, solar metallicity stars (2000 K 
$\le$ T$_{eff}$ $\le$ 4000 K) indicate that the square-root law best
describes the intensity distribution in the infrared.  We ran test 
models of stars with equal temperature and gravity, and found that 
the logarithmic and square-root laws produced nearly indistinguishable 
light curves.  For the final models presented here, the square-root limb
darkening law was adopted.

As shown by Alencar \& Vaz (1999) the limb darkening coefficients of 
stars in close binaries can be effected by irradiation.  Even though 
there is no light coming from the primary object in an SXT, irradiation 
by the accretion disk and hot spot might exist.  While CMCMR did not detect 
a hot spot in the GU Mus system, they did 
see a very strong signature from the secondary star.  They attribute 
this to a very low mass transfer rate.  They also estimated that the 
secondary star contributes 85 - 88\% of the system flux in the 
$R$ band.  Since the K4V secondary will dominate the system flux to a greater 
degree in the infrared, we have used the normal, non-irradiated, limb 
darkening coefficients. We present additional evidence for this in section 3.7.

\subsection{Gravity Darkening}

The second most important static parameter that affects the amplitude
of the ellipsoidal variations is gravity darkening.  Gravity darkening 
(a.k.a. brightening) deals with the localized temperature of a star's 
surface.  The functional form for gravity darkening,
T$_{eff}$ $\propto$ g$^{\beta}$, is wavelength independent, and is 
not strongly affected by changes in either the mixing length or the 
composition of the star.  The amount of gravity darkening depends
on how energy is transported through the star, and thus is correlated 
with the mass of the star.  For low mass, convective stars, $\beta\approx$ 0.32,
 and for stars with radiative envelopes, $\beta \sim$ 1 (Lucy 1967).  For 
the cool secondary star used in our models, we have used a value of 
$\beta$ = 0.38 according to Claret (2000).  He has calculated gravity darkening 
exponents for stars of mass 0.08 M$_{\odot}$ to 40 M$_{\odot}$, and has also 
found that a more accurate method of calculating  gravity darkening exponents 
should include the shape of the star.  Until that is done, however, we will use 
the gravity darkening exponent found by Claret (2000) for a K4V secondary with
T$_{eff}$ = 4,500 K.  This gravity darkening exponent changes very little over 
the estimated spectral type range for the secondary star in GU Mus (K3V $-$ K5V).
                        
\subsection{Other input parameters}

There are a variety of other input parameters with less freedom in their
selection. For example, we must choose the values for the temperature 
and monochromatic luminosities of the secondary star.  For our models, we used 
a K4V secondary temperature of T$_{eff}$=4,500 K, a $J$ band luminosity of L$_J$ 
= 0.311 L$_\odot$, and a $K_s$ band luminosity of L$_{Ks}$ = 0.421 L$_\odot$.

The atmospheres of cool stars are fairly complicated and the details of 
their spectral energy distributions and any changes in such, as a 
function of temperature, make for complex modeling (cf., Allard et al. 
1997). Stellar atmosphere codes have to take into account numerous atomic 
and molecular absorption features which affect the limb darkening 
coefficients.  In order to accurately model the limb darkening effects 
of the secondary star in GU Mus, we have used the Kurucz atmosphere models 
incorporated into WD98. 

Irradiation of the secondary star atmosphere in close binaries such as GU Mus
 can be very important.  Harlaftis \& Filippenko (2000) used doppler tomography
 to show evidence of a hot spot in QZ Vul (=GS 2000+25) where the accretion 
stream meets the accretion disk. WD98 can calculate reflection/re-radiation of 
this type of irradiation based on the bolometric albedos of the two stars.  
The expected 
value for radiative envelopes is unity, while the bolometric albedo for 
the convective secondary star 
is expected to lie somewhere between 0.5 and 1, based on models run by 
Nordlund and Vaz (Nordlund \& Vaz 1990; Vaz \& Nordlund 1985).  This value 
is dependent on the amount of convection in the star: the smaller the mixing 
length parameter, $\alpha$ = $l/H_p$, the closer it is to a radiative 
atmosphere, and the higher the bolometric albedo.  In the case of GU Mus,
we used a bolometric albedo of 0.676 based on the average of the albedos given
 in Table 3 of Nordlund \& Vaz (1990); however, this parameter was irrelevant 
since there is little evidence for other significant sources of luminosity in 
the system. 
                                                           
Our GU Mus $J$ and $K_s$ band observations and their corresponding WD98 
models for five different values of the orbital inclination 
angle (50$^o$, 53$^o$, 55$^o$, 57$^o$, and 60$^o$), are presented in 
Figures 3 and 4, respectively.  The 
points represent the data while the curves represent the models.
We present five different inclinations to demonstrate how a change in 
inclination angle affects the resulting light curve in these two band passes. 
For the models shown in Figs 3 \& 4, a $\chi^2$ test gives $i$ = 55$^o$ as the best fit to the data.  For a fixed set of system parameters, we have found that $\chi^2$ tests can distinguish between orbital inclinations of $\pm$1$^o$.
         
\subsection{Sensitivity of the light curve shape to input parameter variation}

We have just shown how we arrived at our input 
parameters.  Many of the parameters discussed have simply been set to values 
consistent with the system parameters of GU Mus.  Several of these input 
parameters, however, are either derived from observations of GU Mus (T$_{eff}$),
 or are unknown (e.g., the orbital inclination or mass ratio).  In order to 
quantify the sensitivity of the models to the variations of these parameters, 
we ran models with inputs covering a wide range of parameter space.   

Varying T$_{eff}$ from 4,400 K (K5V) to 4,800 K (K3V), the observed spectral 
type range of the secondary star, did not produce significant differences in 
the light curves.  Thus for this entire range of temperatures, with $q$ = 0.10, 
the best fit orbital inclination angle is $i$ = 53$^o$. Since we did not know 
the exact mass of either the primary or secondary star in 
GU Mus, the input values of $q$ had to be estimated from 
previous results. To test the sensitivity of the light curve shape to the mass 
ratio, we ran models with $q$'s covering a very large range: 0.08 $\leq q \leq$ 
0.13.  $\chi^2$ minimization tests were performed on all of these models to 
determine the best fit orbital inclination angle.  For models with a mass ratio 
of $q$ = 0.13, we found a best fit inclination angle of $i$ = 55$^o$.  For 
$q \leq$ 0.1, we found that $i$ = 53$^o$ gave the smallest $\chi^2$ value.  
Thus, our resulting inclination angle will not be very sensitive to $q$. 

\subsection{Sensitivity of the light curve shape to other sources of luminosity in the system}

In the previous sections, we have only considered the light from the secondary star.  If there are other sources of luminosity in the system, they can affect our result for the orbital inclination angle.  For example, excess light from a constant source other than the secondary star will dilute the ellipsoidal variations.  Before we proceed any further, we need to investigate the possibility that other sources of luminosity exist in this system.

As discussed above, CMCMR did not detect a hot spot. They did, however,  see evidence for emission from a disk. If the accretion disk in GU Mus were to contribute infrared light to the system, the true orbital inclination angle would be larger than that derived due to the dilution of the light from the secondary star. CMCMR suggested that there was a small (12 - 15\%) non-stellar contribution to the GU Mus flux in the $R$ band. If we consider a worst case scenario where the contamination at $J$ is the same as that observed at $R$ (15\%), the derived orbital inclination angle is increased by four degrees. Due to the relatively flat nature of model accretion disk spectra (e.g, $f_{\lambda} \propto \lambda^{-2}$; Oke 1977), the contamination will almost certainly be less at $J$ than at $R$.  Assuming this 15\% level of contamination and taking into account the range of mass ratios given in the previous section, we obtain an orbital inclination angle of 54$^o$$^{+4}_{-1.5}$.

We now present evidence which suggests that the level of contamination in the GU Mus system is significantly smaller than stated above.  All of the previous optical studies clearly show unequal maxima in the observed light curves of GU Mus (OBMR, for example).  This implies that one side of the star is heated by the accretion disk, and that contamination is an important factor in the modeling of the light curve.  The observed infrared light curves presented here have equal maxima, and thus we do not see evidence for accretion disk irradiation of the secondary star.  This suggests a minimal level of heating and/or reflection in the GU Mus system.

We can also investigate the level of phase-dependent contamination in GU Mus by noting any changes in the apparent color of the system throughout its orbit.  The infrared color at maximum phase will differ from that at minimum phase if either a hotter or cooler source of luminosity exists within the system.  As can be seen in Figures 3 \& 4, the $J$ $-$ $K$ colors agree at both maximum and minimum phase, thus again ruling out a significant phase-dependent contamination. 

Given that the expected spectral energy distribution of an SXT at minimum light would be more strongly contaminated in the optical portion of the spectrum, we can further examine the level of contamination by looking at the optical colors of GU Mus. In Table 1, we present the colors of GU Mus and compare the averages to the colors of a K4 dwarf and giant (reddened by A$_{\rm V}$ = 0.9 mag). The colors of GU Mus are intermediate to those of the K4 giant and dwarf colors.  Thus, there is very little evidence for significant blue sources of light in the system.  The $V$ $-$ $R$ color is slightly redder ($\sim 10\%$) than the dwarf or giant, but can be explained by emission from H$\alpha$.

Thus, we do not see significant evidence for emission from the accretion disk itself, or its irradiation of the secondary star.  But even in the most conservative scenario, the effect on our derived orbital inclination angle is $\leq$ 4$^o$.

\section{Results}

We have explored a wide range of parameter space for GU Mus, and have found that
 the model light curves are relatively insensitive to the input 
parameters, except for the inclination angle.  Even changing the mass ratio, $q$, by a 
very large amount ($i.e.$, changing the secondary star mass from 0.56 M$_\odot$ 
to 0.90 M$_\odot$ for a primary mass of 6.95 M$_\odot$), changes the derived orbital 
inclination angle of the system by only two degrees ($\pm$1$^o$).  We have considered other sources of luminosity in the system and even under the most conservative scenario, the derived orbital inclination angle is 54$^o$$^{+4}_{-1.5}$.  We believe, however, that the infrared light curve is not seriously affected by any such contamination.  {\it Within the ability of WD98 to realistically model the light curves of SXTs, the derived orbital inclination of GU Mus is 54$\pm$1.5$^o$.}  Our derived orbital inclination angle 
agrees with SNC's inclination determined from their $H$ band light curve of GU Mus.  

With the orbital inclination angle determined, we can estimate the mass of the 
primary. To calculate the mass of the primary object in GU Mus, we need to know 
three things: the orbital inclination, the radial velocity of the secondary 
star, and the mass of the secondary.  The radial velocity of the secondary star 
is K$_2$ = 406$\pm$7 km s$^{-1}$ (OBMR). From the observed range in 
spectral types, the main sequence mass of the secondary star is 0.75$\pm$0.05 
M$_\odot$ (Baraffe et al. 1998). (A large uncertainty in finding the 
mass of the primary remains that associated with the mass of the secondary star.
 As noted by both CMCMR and OBMR, the secondary star in this system may be 
slightly evolved.  Even though the mass of the secondary in GU Mus did not have 
a significant effect on our orbital inclination angle determination, it plays a 
large roll in finding the mass of the primary.) Using these values, we 
calculate a black hole mass of 6.95$\pm$0.6 M$_\odot$.  This mass is within the 
range reported by both CMCMR and SNC.  

For an average primary mass of 6.95 M$_\odot$, and an average secondary mass of 
0.75 M$_\odot$, we obtain a $q$ = 0.108.  Using this mass ratio and a 
semi-major axis of 4.6 R$_\odot$, we find the resulting Roche lobe radius 
of the secondary star in GU Mus to be 0.97 R$_\odot$.  Note that a main sequence K4V would have a radius of 0.73 R$_\odot$, thus filling $\sim$ 75\% of the derived Roche lobe radius.  Interestingly, both the derived radius and optical colors are consistent with the secondary star being a subgiant.  Using R = 0.97 R$_\odot$ and T$_{eff}$ = 4,500 K 
(appropriate for a K4V), we derive a bolometric luminosity of L$_{Bol}$ = 0.34 
L$_\odot$.  Using our infrared photometry, corrected for reddening, we determine a distance to GU Mus of 5.1 kpc. 

The development of a consistent predictive theory of accretion flows around 
black holes is one of the main goals of high-energy astrophysics.  The origin of the outbursts of SXTs can be explained by the accretion disk limit cycle mechanism that was developed to describe the outbursts of cataclysmic variables (Cannizzo 1998). Modeling of the outbursts of SXTs, from the x-ray through optical, has been attempted by a number of researchers.  The consensus appears to be that the optical and soft x-ray luminosities are generated within an accretion disk.  The hard x-ray spectra that develop later in the outburst have been explained as advection dominated accretion onto the black hole.  Depending on the model chosen to explain the multi-wavelength luminosities, the inclination angle can play an important role in the appearance of the outburst.

For example, Esin, McClintock, and Narayan (1997) calculated theoretical x-ray models to fit to various phases that GU Mus exhibited during its outburst.  They combined a normal accretion disk with an advection dominated accretion flow (ADAF).  In their models, they input a black hole mass of 6 M$_\odot$, an inclination angle of $i$ = 60$^o$, and a distance of 5 kpc, taken from OBMR, CMCMR, and SNC.  They found that the during the outburst of GU Mus, both the accretion disk and ADAF remained optically thin. If so, the inclination angle does not play a serious role in what is observed.  But as Esin et al. (2000) note, even if the ADAF is optically thin, the emission will preferentially escape along the polar axes, and thus the inclination angle may affect the observed x-ray flux. In contrast, Misra (1999), uses a similar model (with identical values of M$_1$, $i$, and distance), but one where both the ADAF and the accretion disk are optically thick.  If this model is correct, then the inclination angle plays an important role in what is observed.  New models that are constructed to explain the outburst of GU Mus should take into account our new results for the mass of the black hole, the orbital inclination angle and the distance.

\section{Conclusions}

We have obtained new $J$ and $K_s$ band light curves covering a complete orbit for
 GU Mus.  We have used WD98 to model the resulting light curves, and obtained
 an orbital inclination of 54$\pm$1.5$^o$. This leads to a primary mass estimate 
of 6.95$\pm$0.6 M$_\odot$.  In quiescence, GU Mus is a relatively simple system 
to model: 
 There is no hot spot, and very little accretion to contaminate its infrared 
light curves. There is also a consensus on the orbital inclination angle for GU 
Mus, unlike the other SXTs. Thus, it provides the opportunity to test our 
modeling procedure before attempting more complex systems.   

Before we can fully understand the outbursts of SXTs, we need to determine values for the most important systemic parameters. Unfortunately, it is much 
harder to determine system parameters for non-eclipsing binaries, than for their 
eclipsing counterparts.  The {\it only}
 way to determine the orbital inclination in a non-eclipsing system is to model
 its ellipsoidal variations.  WD98 allows us to model these variations using the most up-to-date software package available. By modeling infrared data, where the contamination from non-stellar sources is minimized, we have a robust technique to determine important orbital parameters. 

\acknowledgments

DMG would like to thank Josef Kallrath and Bob Wilson 
for the use of WD98, as well as their help in running and understanding the 
program.   This research was supported by a Grant-in-Aid of Research from the 
National Academy of Sciences, through Sigma Xi,
The Scientific Research Society. DMG holds an American fellowship from the 
American Association of University Women Educational Foundation. TEH wishes to acknowledge support from the Small Research Grants program of the
American Astronomical Society, the New Mexico Space Grant Consortium, and the New Mexico State University Arts and Sciences Minigrant program.  We would also like to thank the anonymous referee for some useful comments which improved this paper.

\clearpage

\begin{figure}
\figurenum{1}
\plotone{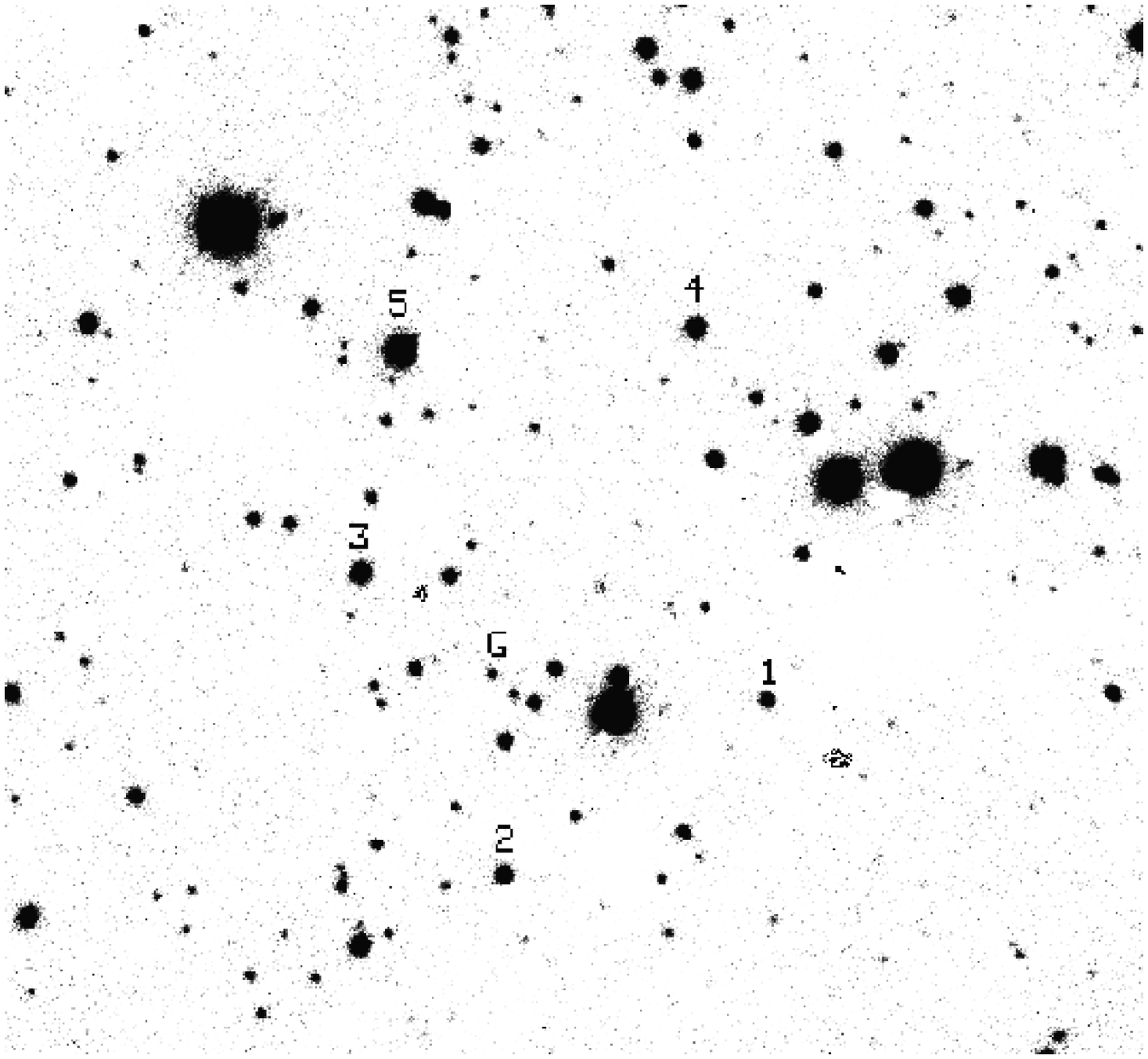}
\caption{OSIRIS $J$ band image of GU Mus ($\alpha_{2000}$ = 11:26:26.7, 
$\delta_{2000}$ = -68:40:32.6) taken on February 20, 2000 with the Victor M. 
Blanco 4 meter telescope at Cerro Tololo Inter-American Observatory. The field
is 1.5' x 1.5' with a scale of 0.16''/pixel.  The SXT is labeled with a ``G''.  
The five comparison stars used for the differential photometry are numbered.}
\end{figure}
         
\begin{figure}
\figurenum{2}
\plotone{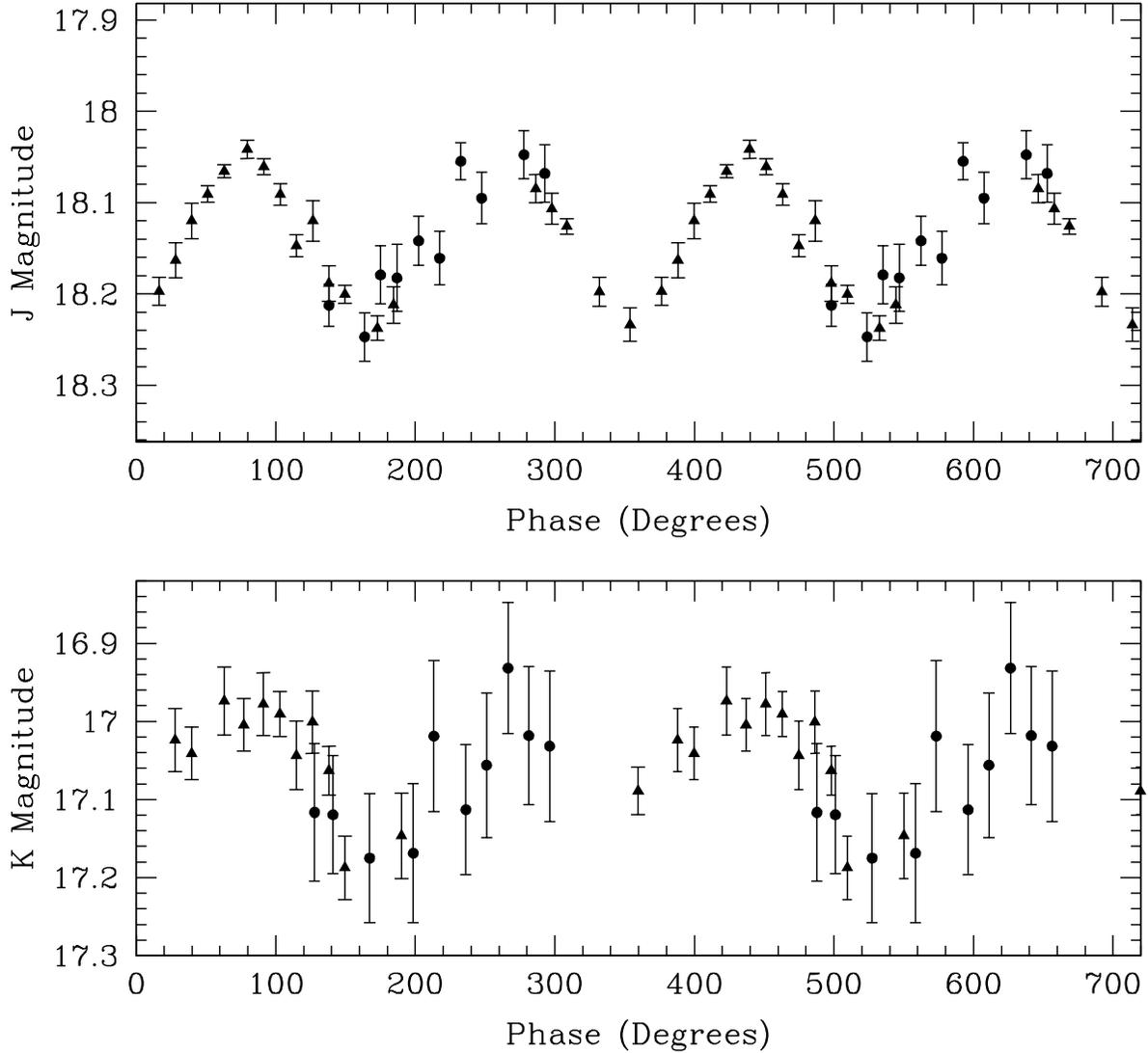}
\caption{GU Mus $J$ band (top panel) and $K_s$ band (bottom panel) light 
curves. The data were obtained on February 20 (triangles) and February 21 
(circles), with OSIRIS on the CTIO 
Victor M. Blanco 4-m telescope.  The data are plotted over two phase cycles 
for clarity.  Here and throughout this paper we phase our 
heliocentric corrected data to the Shahbaz et al. (1997) ephemeris.}
\end{figure} 

\begin{figure}
\figurenum{3}
\plotone{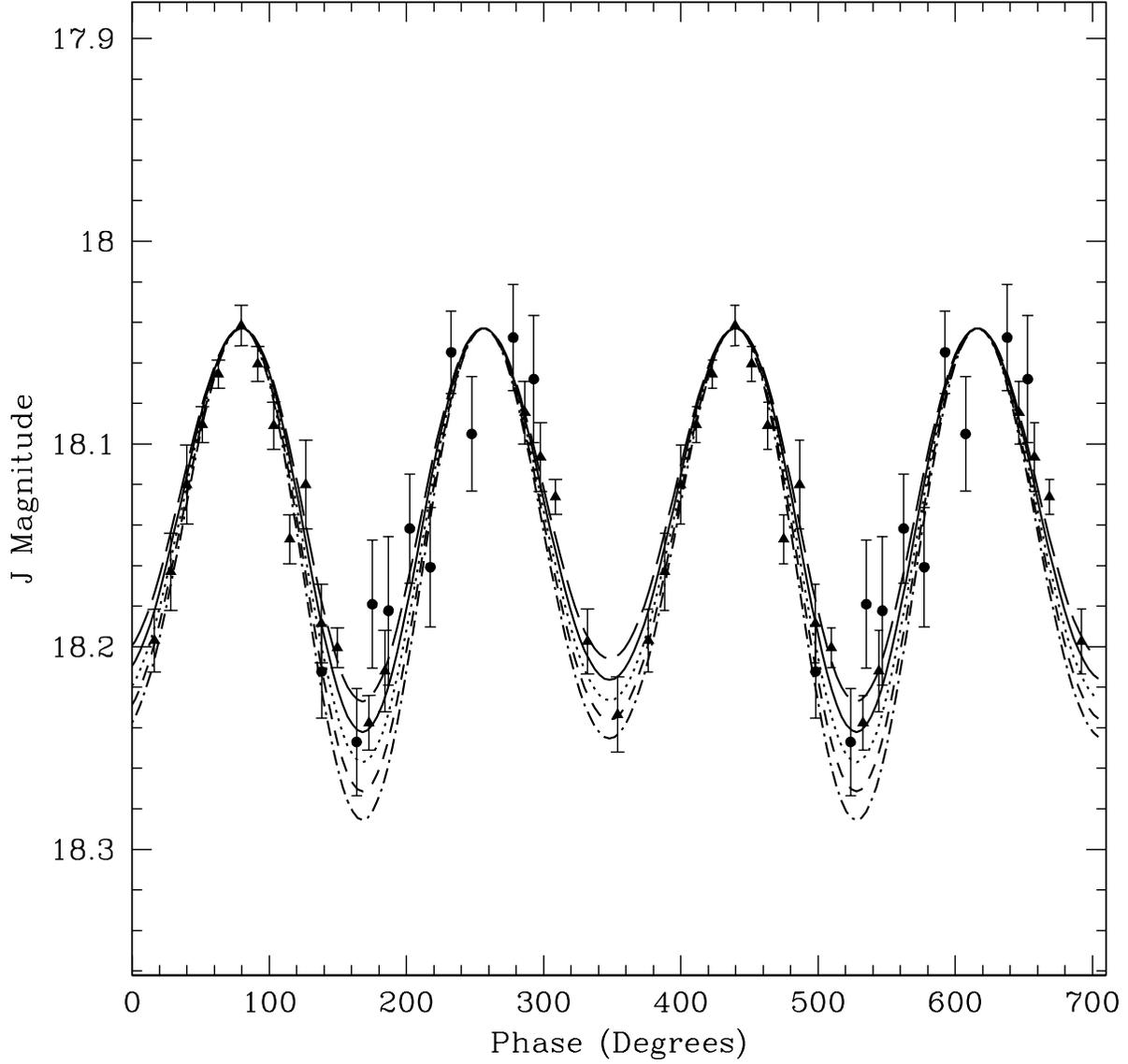}
\caption{$J$ band light curve with five WD98 models.  The data are the same as in
 Figure 2.  The lines are WD98 models of differing orbital 
inclination angles: 50$^o$ (long-dashed), 53$^o$ (solid), 55$^o$ (dotted), 
57$^o$ (short-dashed), and 60$^o$ (dot dashed).  With $q$ = 0.13, a $\chi^2$ minimization finds
 the best fit model to be 55$^o$.}
\end{figure}

\begin{figure}
\figurenum{4}
\plotone{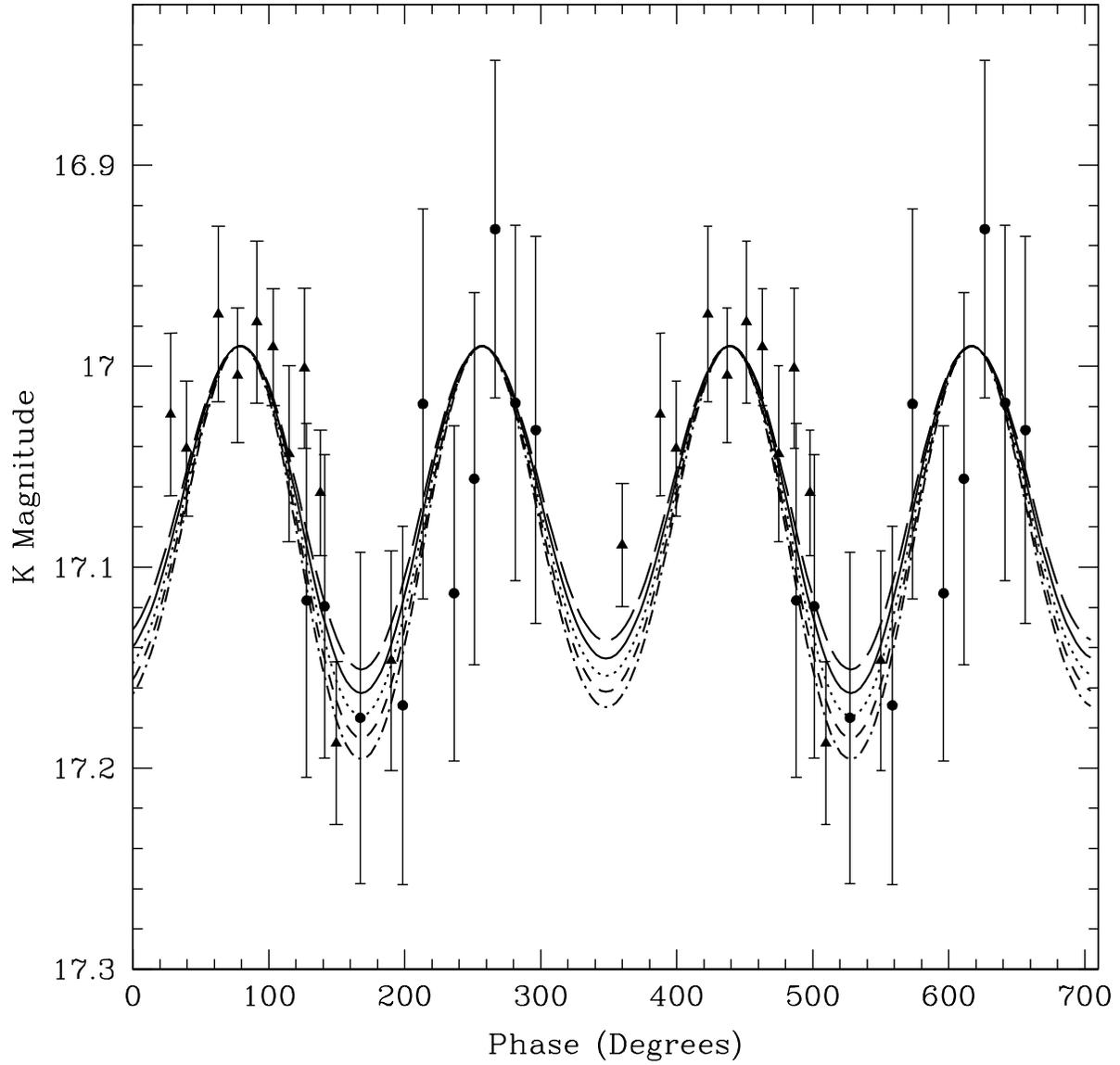}\caption{$K_s$ band light curve with the five WD98 models as in Figure 3.}  
\end{figure}

\clearpage
 
\begin{deluxetable}{cccccccc}
\tabletypesize{\footnotesize}
\tablenum{1}
\tablecaption{Quiescent Infrared and Optical Colors of GU Mus}
\tablehead{
\colhead{Reference} &\colhead{V} &\colhead{B - V} &\colhead{V - R}
&\colhead{V - I} &\colhead{J} &\colhead{J - K} &\colhead{Year\tablenotemark{a}}
}
\startdata
1 & 20.5$\pm$0.1 & 1.60 & 1.00 & ... & ... & ... & 1992 \\
2 & 20.66$\pm$0.03 & ... & 0.91 & 1.66 & ... & ... & 1992 \\
3 & 20.55$\pm$0.05 & ... & ... & 1.53 & ... & ... & 1992 \\
4 & 20.35$\pm$0.03 & ... & ... & 1.55 & ... & ... & 1993 \\
5 & 20.83$\pm$0.06 & ... & 0.90$\pm$0.07 & 1.83$\pm$0.07 & 18.14$\pm$0.02 & 1.06$\pm$0.10 & 2000/2001 \\
&&&&&&& \\
Mean Observed Color & & 1.60$\pm$0.06 & 0.93$\pm$0.05 & 1.64$\pm$0.14 & & 1.06$\pm$0.10 \\
K4V (A$_{\rm V}$ = 0.9 mag) & & 1.36 & 0.77 & 1.53 & & 0.83 \\
K4III (A$_{\rm V}$ = 0.9 mag) & & 1.70 & 0.88 & 1.87 & & 1.03 \\
\enddata
\tablenotetext{a}{Year the data were taken}
\tablerefs{(1) Della Valle et al. 1998; (2) King et al. 1996; (3) Remillard et al. 1992; (4) OBMR; (5) This Paper}
\end{deluxetable}{}

\clearpage                                                      

\begin{deluxetable}{lc}
\tablenum{2}
\tablecaption{Wavelength Independent WD98 Input Parameters for GU Mus}
\tablehead{
\colhead{Parameter}
 &\colhead{Value}
}
\startdata
Orbital Period (days) & 0.432604 \\
Ephemeris\tablenotemark{a}~ (HJD phase 0.0) & 2448812.669 \\
Semi-Major Axis (R$_{\odot}$) & 4.6 \\
Orbital Eccentricity & 0.0 \\
Temperature of K4V Secondary (K) & 4,500 \\                 
Mass Ratio (M$_2$/M$_1$) & 0.108 \\
Atmosphere Model &  Kurucz \\
Limb Darkening Law & Square-root \\ 
Secondary Star Gravity Darkening Exponent & $\beta$=0.38 \\ 
Secondary Star Bolometric Albedo & 0.676 \\
\enddata
\tablenotetext{a} {From Shahbaz, Naylor, and Charles (1997)}
\end{deluxetable}{}

\clearpage

\begin{deluxetable}{ccc}
\tablenum{3}
\tablecaption{Wavelength Dependent WD98 Input Parameters for GU Mus}
\tablehead{
\colhead{Parameter}
 &\colhead{J}
 &\colhead{K$_s$}
}
\startdata
Secondary Star Monochromatic Luminosity (L$_{\odot}$) & 0.311 & 0.421 \\
Secondary Star Square-root Limb Darkening Coefficients: & & \\
x$_\lambda$ & 0.110 & -0.116 \\
y$_\lambda$ & 0.531 & 0.724 \\
\enddata
\end{deluxetable}{}


\newpage

\begin{references}

Allard, F., Huaschildt, P., Alexander, D., \& Starrfield, S. 1997, ARA\&A, 35, 137

Alencar, S.H.P., \& Vaz, L.P.R. 1999, A\&A Supp., 135, 555

Antokhina, Eh. A., \& Cherepashchuk, A. M. 1993, Pis'ma v Astronomicheskij Zhurnal, 19, \#6, 500

Baraffe, I., Chabrier, G., Allard, F., \& Hauschildt, P.H. A\&A, 1998, 337, 403

Bessell, M.S. 1991, AJ, 101, 662

Cannizzo, J. K. 1998, in ASP Conf. Ser., 137, ed. S. Howell, E. Kuulkers, \& C. Woodward (San Francisco: ASP), 308

Casares, J., Mart\'{\i}n, E.L., Charles, P.A., Molaro, P., \& Rebolo, R. 1997, NewA, 1, 299 (CMCMR)

Claret, A. 1998, A\&A, 335, 647

Claret, A. 2000, A\&A, 359, 289

Della Valle, M., Masetti, N., \& Bianchini, A. 1998, A\&A, 329, 606

Di\'az-Cordov\'es, J., \& Gim\'enez, A. 1992, A\&A, 259, 227

Esin, A. A., Kuulkers, E., McClintock, J. E., \& Narayan, R. 2000, ApJ, 532, 1069

Esin, A. A., McClintock, J. E., \& Narayan, R. 1997, ApJ, 489, 865

Harlaftis, E.T., \& Filippenko, A.V. 2000, in Proc. SPIE, 4005, Discoveries and Research Prospects from 8 to 10 - Meter Class Telescopes, ed. J. Bergeron, (New York: Springer), 232

Hua, X., \& Lingenfelter, R.E. 1993, ApJ, 416, L17

Kallrath, J., Milone, E.F. 1999, Eclipsing Binary Stars: Modeling and Analysis (New York: Springer)

Kallrath, J., Milone, E.F., Terrell, D., \& Young, A.T. 1998, ApJ, 508, 308

King, N. L., Harrison, T. E., \& McNamara, B. J. 1996, AJ, 111, 1675

Klinglesmith, D.A., \& Sobieski, S. 1970, AJ, 75, 175

Kurucz, R.L., 1993, in Light Curve Modeling of Eclipsing Binary Stars, ed. E.F. Milone (New York: Springer), 93

Lucy, L.B. 1967, Zeitschr f\"ur Astrophys, 65, 89

Lund, H., Brandt, S. 1991, IAU Circ. 5161

Makino, F. et al. 1991, IAU Circ. 5161

Milne, E.A. 1921, MNRAS, 81, 361

Milone, E.F., Schiller, S.J., Munari, U., \& Kallrath, J. 2000, ApJ, 199, 1405

Misra, R. 1999, ApJ, 512, 340

Nordlund, \AA~., \& Vaz, L.P.R. 1990, A\&A, 228, 231

Oke, J.B. 1977, ApJ, 217, 182

Orosz, J.A., Bailyn, C.D., Remillard, R.A., McClintock, J.E., \& Foltz, C.B. 1994, ApJ, 436, 848

Orosz, J.A., Bailyn, C.D., McClintock, J.E., \& Remillard, R.A. 1996, ApJ, 468, 380 (OBMR)

Pogge, R.W., Martini, P., \& DePoy, D.L. 1999, \\ \url{http://www.ctio.noao.edu/instruments/ir\_instruments/osiris/manual/}

Remillard, R.A., McClintock, J.E., \& Bailyn, C.D. 1992, ApJ, 399, L145

Remillard, R.A., Schachter, J.F., Silber, A.D., \& Slane, P. 1994, ApJ, 426,
288

Shahbaz, T., Naylor, T., \& Charles, P.A. 1997, MNRAS, 285, 607 (SNC)

van Hamme, W. 1993, AJ, 106, 2096

Vaz, L.P.R., \& Nordlund, \AA~. 1985, A\&A, 147, 281

Wilson, R.E. 1979, ApJ, 234, 1054

Wilson, R.E. 1990, ApJ, 356, 613

Wilson, R.E. 1993, in ASP Conf. Ser. 38, New Frontiers in Binary Star Research, ed. K.C. Leung \& I.S. Nha (San Francisco: ASP), 91

Wilson, R.E. 1998, in Reference Manual to the Wilson-Devinney Program, Computing Binary Star Observables, Version 1998 (Gainsville, FL: Univ. Florida)

Wilson, R.E., \& Devinney, E.J. 1971, ApJ, 166, 605

\end{references}
\end{document}